**Accepted. Appl. Phys. A (2022)**

# Investigation of structural and magneto-transport properties of $PdTe_2$ single crystals

Yogesh Kumar[1,2,3], Prince Sharma[1,2,4], M.M. Sharma[1,2] and V.P.S. Awana[1,2,*]

[1]*CSIR-National Physical Laboratory, Dr. K. S. Krishnan Marg, New Delhi-110012, India*

[2]*Academy of Scientific and Innovative Research (AcSIR), Ghaziabad 201002, India*

[3]*School of Science, RMIT University, Melbourne, VIC 3001, Australia*

[4]*Functional Materials and Microsystems Research Group and the Micro Nano Research Facility, RMIT University, Melbourne, VIC, 3001, Australia*

## Abstract

The growth and structural analysis of topological semimetal (TSM) $PdTe_2$ were carried out in this study. The self-flux method was employed to grow the single crystals which were structurally characterized by XRD, SEM and EDAX, while the vibrational modes were observed by Raman spectroscopy. Further, the transport properties of the grown crystal are also investigated, which show the presence of the weak anti-localization effect. The low field (≈1Tesla) magnetoconductivity is studied by Hikami-Larkin-Nagaoka (HLN) model, and the physical parameters are extracted. Further, a quadratic term and a linear term in applied field were added in HLN model, which are accounted for quantum and classical contributions in conduction mechanism. The modified HLN model was used to study magneto-conductivity in entire field range and at temperature from 2-100 K. This study not only confirmed the growth of single crystal $PdTe_2$, but also verified the presence of topological surface states (TSS) through the HLN analysis of magneto-transport measurements.

***Corresponding Author**
Dr. V. P. S. Awana: E-mail: awana@nplindia.org
Ph. +91-11-45609357, Fax-+91-11-45609310
Homepage: awanavps.webs.com

## Introduction

Quantum materials such as transition metal dichalcogenides (TMDs) with topological non-trivial surface states (TSS) are currently drowning much attention of condensed matter scientists as it offers an ideal platform for experimental realization of the Dirac fermions [1-6]. The quantum materials include topological insulators (TIs), topological semimetals (TSMs), and topological superconductors (TSCs) [7-12]. These materials show numerous applications, including thermoelectricity, terahertz generation, superconductivity, quantum computing, and optoelectronic [13-20]. While especially coming to TSMs, they show exotic properties, such as large magnetoresistance (MR) [21], non-ohmic transport [22], and negative MR [1]. Dirac Semimetal, Weyl Semimetal, and Nodal lines Semimetal are the classes of TSMs. While, the TSMs can be classified into two more classes, viz. Type-I TSM and Type-II TSM according to their band topology. Band dispersion in type-I TSM is linear cone-shaped as that of $Cd_3As_2$ [23] and TaP [24], and breaking Lorentz invariance in some special TSMs forms another class of TSMs, i.e., type-II TSMs. The Dirac/ Weyl cone is inclined along a specific momentum direction, and these Dirac/Weyl points appear as just contact between electron and hole pockets [5]. Due to this unique band topology, type-II TSMs possess many exotic properties, such as direction-dependent chiral anomaly, unusual superconductivity and unusual quantum oscillations [25-27].

Materials that can be regarded as type-II TSMs include some Transition Metal Dichalcogenides (TMDs); TMDs are 2-D layered materials in which atomic layers are separated by a Vander Waals gap, making them a good candidate for material intercalation. TMDs like $PtTe_2$ [28], $PtSe_2$ [4,28], $PdTe_2$ [3], $NiTe_2$ [29] are considered to be type-II Dirac semimetals. $PdTe_2$ is found to be superconducting at 1.7K [30]**,** which can be increased up to 4.7K by Au intercalation [31]. It makes $PdTe_2$ to be an encouraging material to realize topological superconductivity. The observed MR in TSMs is not linear and is found to increase in a quadratic manner [32] as the transport in these materials is dominated by non-relativistic carriers because the Dirac point in these materials lies far below the Fermi level [28]. These intrinsic noble characteristics of TSMs, including the topological superconductivity and non-linear MR, increase the eagerness to study and probe the topological surface states.

In this paper, the growth, structural analysis and magneto-transport measurements of $PdTe_2$ single crystals are carried out. A well optimized heat treatment based on the self-flux method is followed to grow the single crystals. The as grown crystals are characterized through various techniques viz. X-ray diffraction (XRD), Scanning electron microscopy (SEM), and energy-dispersive X-ray spectroscopy (EDAX) and Raman spectroscopy. The transport properties show the presence of the WAL effect (weak antilocalisation) and thereby topological character at low temperature in as grown $PdTe_2$ single crystal. Further, the modified HLN is also used at higher fields to study the quantum and classical contributions in magneto-conductivity.

## Experimental details

Single crystal $PdTe_2$ was grown using a self-flux method, in which high purity (4N, Sigma-Aldrich) powders of Pd and Te were taken in stoichiometric ratio and grounded thoroughly by using an agate mortar and pestle. To avoid any oxidation during this process, grinding of powders was performed in MBRAUN glove box filled with Ar gas. The finely grounded powder mixture was then palletized using a hydraulic press, and thus obtained pallet was encapsulated in a quartz tube. The vacuum level inside the quartz tube was maintained at an order of $10^{-5}$ mbar using a diffusion pump. This encapsulated sample was placed in a (Proportional-Integral-Derivative) PID-controlled Muffle furnace, and heated to 850°C at a heating rate of 120°C/h followed by a 24 h hold in order to obtain a homogeneous molten mixture. Then, the sample is slowly cooled down to 550°C at a rate of 1°C/h. Through this slow cooling, the layer-by-layer growth of $PdTe_2$ takes place. Then this sample was annealed at 550°C for 48 h, and the furnace was switched off in order to allow them to cool down to room temperature. The obtained crystal was silvery shiny and easily cleavable along its growth axis. The schematic of the optimized heat treatment diagram for single crystal growth of $PdTe_2$ is shown in fig.1.

The XRD spectra were recorded to determine the phase of grown crystals using a Rigaku made table-top X-Ray Diffractometer equipped with Cu-$K_\alpha$ radiation of 1.5418 Å. The XRD patterns were recorded for both viz. mechanically cleaved crystal flake and gently crushed powder of grown crystal. The powder XRD pattern is fitted using FullProf Software, and the unit cell structure is drawn using VESTA software. The composition and layered geometry of the crystal were confirmed by scanning electron microscopy (SEM, model: Zeiss EVOMA 10) and elemental analysis performed by energy-dispersive X-ray spectroscopy (EDAX, model: Oxford INCA). To study the vibrational modes of grown crystal, Raman spectrum was recorded using Renishaw inVia Reflex Raman Microscope equipped with a laser working at 532 nm. Mechanically cleaved flat crystal flake was irradiated at 532 nm for 10 s at a focus of L50X. Laser power is maintained below 5 mW during the irradiation process to avoid any local heating of the flake surface. A Quantum Design physical property measurement system (PPMS) is used to study the transport properties, where the RT (Resistance vs. temperature) and RH (resistance vs. magnetic field) curves were recorded at different fields and temperatures.

## Results and Discussion

The powder XRD spectrum were recorded on gently crushed powder of thin flakes of the grown crystal to confirm the phase of the system. Fig.2(a) shows the Rietveld refinement of the powder XRD data of $PdTe_2$ crystals. It confirms that synthesized $PdTe_2$ crystallizes in $CdI_2$ type trigonal structure which belongs to $P\bar{3}m1$ space group. The agreement of Rietveld data with observed data shows that crystals were grown in single phase and impurity peaks are absent in the XRD pattern. The quality of refinement can be estimated by the goodness of fit $\chi^2$, which is found to be 2.96, and is lies in acceptable range. Refined lattice parameters obtained from Rietveld refinement are $a = b = 4.042(22)$ Å and $c = 5.142(52)$ Å, $\alpha = \beta = 90°$ and $\gamma = 120°$ for $PdTe_2$. The inset in fig. 2(a) shows the unit cell structure drawn by using VESTA software in which Pd atoms occupies (0, 0, 0) position and the Te atoms occupy (0.333,0.667,0.25) positions.

Further, in order to investigate the single crystalline growth, thin flakes are mechanically cleaved along their growth axis from as-grown single crystals. Fig. 2(b) shows the XRD spectra recorded on $PdTe_2$ flakes over a 2θ range of 10°-80°. It is observed that the diffraction peaks arise from (001), (002), (003), and (004) planes of $PdTe_2$. The occurrence of high-intensity peaks only for (*00l*) planes confirms the unidirectional crystal growth along the c-axis.

The XRD pattern confirms the phase purity of as-grown crystal, and further, the compositional stoichiometry and the unidirectional growth of the crystals are investigated by EDAX and SEM, respectively. The inset of fig.3 shows the SEM image and the atomic percentage of the constituent elements of grown crystal. The obtained atomic percentage confirms the nearly stoichiometry of $PdTe_2$, and the SEM image confirms the unidirectional growth of the crystal as the flake shows a planar nature. To study the vibrational modes, the crystal flake is further investigated using Raman spectroscopy. Fig. 3 shows recorded Raman spectra of synthesized $PdTe_2$ single crystals. Here, in the measurements, mainly two Raman Shift peaks are observed at 83.8 and 125.6 $cm^{-1}$, which are attributed to $E_g$ and $A_{1g}$ modes of $PdTe_2$ crystal, respectively and these values are in consistent with previous reports on $PdTe_2$ [33,34]. The Raman active $E_g$ modes occur due to in-plane movement of Te atoms in $PdTe_2$ unit cell, while Raman active $A_{1g}$ mode occurs due to out of plane movement of Te atoms, as shown by schematic in fig. 3. Both of these modes are mainly dependent on the motion of Te atoms.

Further, the transport properties of these grown crystals are investigated using QD-PPMS, where the flakes of these crystals are examined, and four-probe geometry is used to carry out the transport measurements. Fig. 4(a) depicts the behaviour of electrical resistivity (ρ) as a function of temperature for single crystal $PdTe_2$ in absence of any applied field. It is observed that resistivity decreases from 130.79 μΩ-cm at 295 K to 7.74 μΩ-cm at 2 K, as the temperature is decreased from 295 down to 2 K. The residual resistivity ratio RRR = ρ (295 K) / ρ (2 K) ≈ 17, indicates the high quality of single crystal. The temperature dependence of electrical resistivity can be explained by Boltzmann transport theory using Bloch-Gruneisen (BG) formula [35], which is given as

$$\rho(T) = \rho(0) + \rho_{el-ph}(T) \tag{1}$$

where the first term $\rho(0)$ is the residual resistivity, which is temperature independent and arises from defect scattering. Next, the temperature dependent part of resistivity is described by second term and this component depends on electron-phonon scattering. $\rho_{el-ph}(T)$ is given by the formula

$$\rho_{el-ph}(T) = \alpha_{el-ph}\left(\frac{T}{\Theta_D}\right)^n \int_0^{\Theta_D/T} \frac{x^n}{(e^x-1)(1-e^{-x})}dx \tag{2}$$

where $\Theta_D$ represents the Debye temperature, $\alpha_{el-ph}$ is the electron-phonon coupling parameter and $n$ is a constant. The solid black curve in fig. 4(a) represents the best fit of measured resistivity for n = 5, denoting the presence of electron-phonon interaction. The fitting by BG formula yields $\rho(0)$ = 8.02 μΩ-cm and $\Theta_D$ = 180.8 K. Further, the electrical resistivity is measured in low temperature regime from 100 to 2 K in the presence of applied transverse magnetic fields 0, 3, 5, and 12 Tesla as shown in inset of fig 4(a). It is observed that resistivity decreases with decreasing temperature at all fields, ensuring the metallic nature. Although, a slight increment in resistivity is observed as the field strength is increased from 0 to 12 Tesla.

The magnetoresistance (MR) of $PdTe_2$ flakes is also investigated by measuring the RH (resistance vs. magnetic field) over a transverse magnetic field range of up to ±12 Tesla at different temperatures. The MR% value is calculated using a general formula where, first, the resistivity of the flakes at the different magnetic fields is examined, and then, MR % is calculated using MR% = $((\rho_H-\rho_0)/\rho_0)$ ×100%. Here, $\rho_0$ and $\rho_H$ are the resistivities at zero and applied magnetic fields, respectively. Fig. 4(b) shows the MR% of $PdTe_2$ single crystal at different temperatures. At a magnetic field of 12 Tesla, the MR% is around 85% for temperature 2 K and this MR is non-saturating in the entire measured field range. Further, this MR% decreased to 6% as the temperatures is decreased to 100 K for magnetic field of 12 Tesla. In addition to non-saturating MR at high field, a V-shaped response is also observed in low magnetic field regime, which corresponds to the presence of WAL (weak anti-localization) effect. The inset of fig. 4(b) represents the variation of MR% with respect to magnetic field up to ± 1 Tesla and indicates the observed cusp like behaviour. The occurrence of WAL effect has been reported in various topological insulator systems.

Further, to study the different carrier contributions and scattering mechanism arising at different temperatures and magnetic fields in conduction, the Kohler rule is applied on magnetoresistance of $PdTe_2$ single crystal. As per the semiclassical transport theory, the Kohler rule holds if there is a contribution from single type of charge carriers at different magnetic fields. According to this rule, the variation in resistivity i.e., $\Delta\rho$ / $\rho_0$ as a function of applied magnetic field can be described as a function of H$\tau$. Here, $\Delta\rho$ is described as the difference between resistivity at applied field ($\rho$(H)) and resistivity at zero field ($\rho$(0)). The variable $\tau$ is inversely proportional to zero field resistivity $\rho_0$ and is defined as the average time between two consecutive scatterings of conduction electrons [36,37]. Fig. 4(c) represents the variation of MR% vs H / $\rho_0$ at different temperatures from 100 to 2 K. It is observed that MR% curves at different temperatures do not fall into a single curve, hence violating the Kohler rule. This deviation corresponds to the presence and contribution from multiple charge carriers at different temperatures in the $PdTe_2$ single crystal.

Now, to investigate the observed WAL effect in low field regime and to extract information regarding conduction channel and phase coherence length, the magneto-conductivity (MC) is analyzed. The conductivity at particular field and temperature is calculated by taking inverse of respective resistivity. Fig. 5(a) represents the observed magneto-conductivity vs magnetic field of up to ±1 Tesla at temperature varied from 100 K down to 2 K. Here, the magnetic field is applied perpendicular to both direction of current flow and sample surface. A sharp V-type cusp is observed at all measured temperatures near low-field regime and also, this broadens with increase in temperature. According to the previous reports on topological insulators [38], the occurrence of negative magneto-conductivity corresponds to the presence of WAL effect in present $PdTe_2$ single crystal. In topological insulators, the parameters describing the low temperature WAL effect has been explained by Hikami-Larkin-Nagaoka (HLN) model [39]. According to HLN model, the quantum correction to 2D magneto-conductivity is given by the equation

$$\Delta\sigma(H) = -\frac{\alpha e^2}{\pi h}\left[ln\left(\frac{B_\varphi}{H}\right) - \Psi\left(\frac{1}{2} + \frac{B_\varphi}{H}\right)\right] \quad (3)$$

here, Ψ is the digamma function, $L_\varphi$ is the phase coherence length, h is the Plank's constant, e is the electronic charge and $B_\varphi = \frac{h}{8e\pi L_\varphi^2}$ is the characteristic field. In this equation, the free fitting parameters are pre-factor $\alpha$ and phase coherence length $L_\varphi$. For each surface conduction channel, the pre-factor $\alpha$ takes the value -0.5, but as per previous reports, the value of $\alpha$ deviates from these values [40,41]. Moreover, in order to evaluate the surface conduction channel and phase coherence length, the observed magneto-conductivity is fitted using HLN model in low field regime (up to ±1 Tesla) at all measured temperatures. Fig. 5(a) represents the magneto-conductivity as a function of transverse magnetic field at different temperatures. Here, the symbols show experimentally observed data and solid curve shows the HLN fitted curve. The pre-factor $\alpha$ and phase coherence length were determined by fitting the magneto-conductivity curve using equation 3. At 2 K, the extracted value of $L_\varphi$ is ≈ 102 nm and value of $\alpha$ is -0.347, which is close to theoretical value of -0.5. The value of pre-factor suggests the presence of single conduction channel and also the presence of WAL effect in grown $PdTe_2$ single crystal. The extracted values of pre-factor $\alpha$ and phase coherence length $L_\varphi$ by applying HLN equation in low field (up to ±1 Tesla) at different temperatures are shown in table 1. It is observed that $\alpha$ value increases with increment in temperature and phase coherence length decreases with increase in temperature. This suggests the weakening of WAL effect as the temperature is increased from 2 to 100 K in the $PdTe_2$ single crystal.

Further, as the low field regime behaviour is well explained by HLN model, but it is also observed that HLN fitted curve deviates from experimental magneto-conductivity at higher fields. This may be arising due to the contribution from quantum scattering, cyclotronic magnetoresistance or bulk contribution to magneto-conductivity, which occurs at higher magnetic fields and temperatures. To study the contribution of various factors in overall conduction, the magneto-conductivity is fitted by adding different terms in conventional HLN model. As per previous reports, a field dependent quadratic term ($\beta H^2$) considers both quantum ($\beta_q$) and classical ($\beta_c$) effect which contribute in conduction mechanism. Here, $\beta_q$ consists of spin-orbit scattering and elastic scattering whereas, $\beta_c$ explains cyclotronic magnetoresistance [42]. Moreover, in addition to quadratic term consisting quantum and classical contributions, a linear term in field ($\gamma H$) considering the effect of temperature in overall conduction mechanism is also added in HLN model. The coefficient of linear term describes the contribution of bulk carriers in magneto-conductivity of material. As per literature, the surface charge carriers contribute in low temperature conduction, whereas the contribution of bulk charge carriers to magneto-conductivity increases with increase in temperature [43,44]. In order to study the effect of these contributions on conduction mechanism of $PdTe_2$ single crystal, the magneto-conductivity is fitted by using the equation HLN + $\beta H^2$ + $\gamma H$. In fig. 5(b), the solid curve represents the fitted magneto-conductivity using above equation over a magnetic field range of ±12 Tesla at all measured temperatures. For the best fit to experimental data, the extracted values of parameters $\alpha$, $L_\varphi$, $\beta$ and $\gamma$ are listed in table 2. It is observed that coefficient $\gamma$ increases with the increase in temperature, and it suggests that the bulk contribution increases with temperature. Thus, the transport property predicts the topological character of $PdTe_2$ single crystal and fitting of the observed data using mentioned equation suggests that low temperature conduction is dominated by surface carriers, whereas, with increasing temperature and field, the bulk carriers start contributing in conduction.

## Conclusions

The $PdTe_2$ single crystal was grown by the solid-state reaction route through self-flux method. The XRD, SEM and EDAX measurements confirm the single crystal growth, phase purity and compositional exactness of grown crystal, while the Raman spectroscopy confirms the presence of Raman-active vibrational modes $E_g$ and $A_{1g}$. Moreover, the transport properties show the presence of the WAL effect in low field regime, which is analysed by applying HLN model, and the extracted physical parameters suggests the presence of single conduction channel at 2 K. The deviation of HLN fitted curve from experimental observed magneto-conductivity is accounted by HLN + $\beta H^2$ + $\gamma H$ equation over a magnetic field range of ±12 Tesla and at all temperatures. The additional quadratic and linear terms explain the contributions from quantum scattering, cyclotronic magnetoresistance and bulk charge carriers as the strength of field and temperature are increased.

## Acknowledgment

The director of NPL strongly supports this work. The authors would like to thank Mr. K.M Kandpal for vacuum encapsulation of samples for heating in furnace. Authors also thanks CSIR and UGC for financial support and AcSIR for enrolment as a research scholar in Ph.D. program.

**References:**

1. S. Wang, B. C. Lin, A. Q. Wang, D. P. Yu, and Z. M. Liao, Adv. Phys. X **2**, 518 (2017).
2. H. J. Noh, J. Jeong, E. J. Cho, K. Kim, B. I. Min, and B. G. Park, Phys. Rev. Lett. **119**, 016401 (2017).
3. F. Fei, X. Bo, R. Wang, B. Wu, J. Jiang, D. Fu, M. Gao, H. Zheng, Y. Chen, X. Wang, H. Bu, F. Song, X. Wan, B. Wang, and G. Wang, Phys. Rev. B **96**, 041201 (2017).
4. K. Zhang, M. Yan, H. Zhang, H. Huang, M. Arita, Z. Sun, W. Duan, Y. Wu, and S. Zhou, Phys. Rev. B **96**, 125102 (2017).
5. A. A. Soluyanov, D. Gresch, Z. Wang, Q. Wu, M. Troyer, X. Dai, and B. A. Bernevig, Nature **527**, 495 (2015).
6. E. Li, R. Z. Zhang, H. Li, C. Liu, G. Li, J. O. Wang, T. Qian, H. Ding, Y. Y. Zhang, S. X. Du, X. Lin, and H. J. Gao, Chinese Phys. B **27**, 086804 (2018).
7. A. Bansil, H. Lin, and T. Das, Rev. Mod. Phys. **88**, 021004 (2016).
8. X. L. Qi and S. C. Zhang, Rev. Mod. Phys. **83**, 1057 (2011).
9. A. Burkov, Nat. Mater. **15**, 1145 (2016).
10. V. V. Atuchin, V. A. Golyashov, K. A. Kokh, I. V. Korolkov, A. S. Kozhukhov, V. N. Kruchinin, S. V. Makarenko, L. D. Pokrovsky, I. P. Prosvirin, K. N. Romanyuk, and O. E. Tereshchenko, Cryst. Growth Des. **11**, 5507 (2011).
11. K. A. Kokh, V. V. Atuchin, T. A. Gavrilova, N. V. Kuratieva, N. V. Pervukhina, and N. V. Surovtsev, Solid State Commun. **177**, 16 (2014).
12. V. V. Atuchin, V. A. Golyashov, K. A. Kokh, I. V. Korolkov, A. S. Kozhukhov, V. N. Kruchinin, I. D. Loshkarev, L. D. Pokrovsky, I. P. Prosvirin, K. N. Romanyuk, and O. E. Tereshchenko, J. Solid State Chem. **236**, 203 (2016).
13. S. Bano, A. Kumar, B. Govind, A. H. Khan, A. Ashok, and D. K. Misra, J. Mater. Sci. Mater. Electron. **31**, 8607 (2020).
14. J. Sánchez-Barriga, E. Golias, A. Varykhalov, J. Braun, L. V. Yashina, R. Schumann, J. Minár, H. Ebert, O. Kornilov, and O. Rader, Phys. Rev. B **93**, 155426 (2016).
15. B. Keimer and J. E. Moore, Nat. Phys. **13**, 1045 (2017).
16. J. Sánchez-Barriga, M. Battiato, M. Krivenkov, E. Golias, A. Varykhalov, A. Romualdi, L. V. Yashina, J. Minár, O. Kornilov, H. Ebert, K. Held, and J. Braun, Phys. Rev. B **95**, 125405 (2017).
17. N. Kumar, B. A. Ruzicka, N. P. Butch, P. Syers, K. Kirshenbaum, J. Paglione, and H. Zhao, Phys. Rev. B **83**, 235306 (2011).

18. K. Shrestha, M. Chou, D. Graf, H. D. Yang, B. Lorenz, and C. W. Chu, Phys. Rev. B **95**, 195113 (2017).
19. J. W. McIver, D. Hsieh, S. G. Drapcho, D. H. Torchinsky, D. R. Gardner, Y. S. Lee, and N. Gedik, Phys. Rev. B - Condens. Matter Mater. Phys. **86**, 035327 (2012).
20. Y. Ando, J. Phys. Soc. Japan 82, 102001 (2013).
21. L. M. Schoop, L. S. Xie, R. Chen, Q. D. Gibson, S. H. Lapidus, I. Kimchi, M. Hirschberger, N. Haldolaarachchige, M. N. Ali, C. A. Belvin, T. Liang, J. B. Neaton, N. P. Ong, A. Vishwanath, and R. J. Cava, Phys. Rev. B - Condens. Matter Mater. Phys. **91**, 214517 (2015).
22. X. Huang, L. Zhao, Y. Long, P. Wang, D. Chen, Z. Yang, H. Liang, M. Xue, H. Weng, Z. Fang, X. Dai, and G. Chen, Phys. Rev. X **5**, 031023 (2015).
23. M. Neupane, S. Y. Xu, R. Sankar, N. Alidoust, G. Bian, C. Liu, I. Belopolski, T. R. Chang, H. T. Jeng, H. Lin, A. Bansil, F. Chou, and M. Z. Hasan, Nat. Commun. **5**, 3786 (2014).
24. S. M. Huang, S. Y. Xu, I. Belopolski, C. C. Lee, G. Chang, B. Wang, N. Alidoust, G. Bian, M. Neupane, C. Zhang, S. Jia, A. Bansil, H. Lin, and M. Z. Hasan, Nat. Commun. **6**, 7373 (2015).
25. Y. Y. Lv, X. Li, B. Bin Zhang, W. Y. Deng, S. H. Yao, Y. B. Chen, J. Zhou, S. T. Zhang, M. H. Lu, L. Zhang, M. Tian, L. Sheng, and Y. F. Chen, Phys. Rev. Lett. **118**, 096603 (2017).
26. M. Alidoust, K. Halterman, and A. A. Zyuzin, Phys. Rev. B **95**, 155124 (2017).
27. T. E. O’Brien, M. Diez, and C. W. J. Beenakker, Phys. Rev. Lett. **116**, 236401 (2016).
28. W. Zheng, R. Schönemann, N. Aryal, Q. Zhou, D. Rhodes, Y. C. Chiu, K. W. Chen, E. Kampert, T. Förster, T. J. Martin, G. T. McCandless, J. Y. Chan, E. Manousakis, and L. Balicas, Phys. Rev. B **97**, 235154 (2018).
29. C. Xu, B. Li, W. Jiao, W. Zhou, B. Qian, R. Sankar, N. D. Zhigadlo, Y. Qi, D. Qian, F. C. Chou, and X. Xu, Chem. Mater. **30**, 4823 (2018).
30. Amit and Y. Singh, Phys. Rev. B **97**, 054515 (2018).
31. K. Kudo, H. Ishii, and M. Nohara, Phys. Rev. B **93**, 140505 (2016).
32. T. Liang, Q. Gibson, M. N. Ali, M. Liu, R. J. Cava, and N. P. Ong, Nat. Mater. **14**, 280 (2015).
33. S. Y. Liu, S. X. Zhu, Q. Y. Wu, C. Zhang, P. B. Song, Y. G. Shi, H. Liu, Z. T. Liu, J. J. Song, F. Y. Wu, Y. Z. Zhao, X. F. Tang, Y. H. Yuan, H. Huang, J. He, H. Y. Liu, Y. X. Duan, and J. Q. Meng, Results Phys. **30**, 104816 (2021).

34. E. Li, R.-Z. Zhang, H. Li, C. Liu, G. Li, J.-O. Wang, T. Qian, H. Ding, Y.-Y. Zhang, S.-X. Du, X. Lin, and H.-J. Gao, Chinese Phys. B 27, 086804 (2018).
35. J. M. Ziman, Electrons and Phonons: The Theory of Transport Phenomena in Solids (Clarendon Press, Oxford), 1960.
36. Y. C. Luo, Y. Y. Lv, R. M. Zhang, L. Xu, Z. A. Zhu, S. H. Yao, J. Zhou, X. X. Xi, Y. B. Chen, and Y. F. Chen, Phys. Rev. B **103**, 064103 (2021).
37. N. H. Jo, Y. Wu, L. L. Wang, P. P. Orth, S. S. Downing, S. Manni, D. Mou, D. D. Johnson, A. Kaminski, S. L. Bud'Ko, and P. C. Canfield, Phys. Rev. B **96**, 165145 (2017).
38. H. Z. Lu and S. Q. Shen, in Spintron. VII, 91672E (2014).
39. S. Hikami, A. I. Larkin, and Y. Nagaoka, Prog. Theor. Phys. **63**, 707 (1980).
40. G. Zhang, H. Qin, J. Chen, X. He, L. Lu, Y. Li, and K. Wu, Adv. Funct. Mater. **21**, 2351 (2011).
41. H. T. He, G. Wang, T. Zhang, I. K. Sou, G. K. L. Wong, J. N. Wang, H. Z. Lu, S. Q. Shen, and F. C. Zhang, Phys. Rev. Lett. **106**, 166805 (2011).
42. B. A. Assaf, T. Cardinal, P. Wei, F. Katmis, J. S. Moodera, and D. Heiman, Appl. Phys. Lett. 102, 012102 (2013).
43. A. Jash, S. Ghosh, A. Bharathi, and S. S. Banerjee, Bull. Mater. Sci. **45**, 17 (2022).
44. A. Jash, S. Ghosh, A. Bharathi, and S. S. Banerjee, Phys. Rev. B **101**, 165119 (2020).

**Table 1:** HLN fitted parameters of $PdTe_2$ single crystals in low field regime (up to ±1Tesla).

| **Temperature (K)** | **α** | **$L_\varphi$ (nm)** |
|---|---|---|
| 2 | -0.347 | 101.99 |
| 20 | -0.272 | 62.79 |
| 30 | -0.154 | 41.93 |
| 100 | -0.014 | 40.62 |

**Table 2:** Parameters extracted from fitting the magneto-conductivity of $PdTe_2$ single crystals using equation $HLN+\beta H^2+\gamma H$.

| **Temperature (K)** | **α** | **$L_\varphi$ (nm)** | **B** | **γ** |
|---|---|---|---|---|
| 2 | -0.339 | 14.95 | $5.7\times10^{-3}$ | -0.254 |
| 20 | -0.267 | 14.74 | $2.1\times10^{-3}$ | -0.162 |
| 30 | -0.146 | 12.45 | $1.6\times10^{-3}$ | -0.048 |
| 100 | -0.013 | 14.78 | $-6.1\times10^{-3}$ | 0.003 |

**Figure captions:**

Fig. 1: Optimized heat treatment for the growth of $PdTe_2$ single crystals.

Fig. 2: (a) Rietveld refinement of powder XRD spectra of $PdTe_2$ and inset shows the unit cell structure shows respective atomic positions. (b) XRD spectra on mechanically cleaved thin flakes of obtained crystals.

Fig. 3: Room temperature Raman spectroscopy shows the obtained vibrational modes for $PdTe_2$ single crystals. Inset shows the SEM images depicting the layered growth and EDAX spectroscopy represents the elemental composition of respective constituents.

Fig. 4: (a) Bloch-Gruneisen fitting to the measured resistivity in zero magnetic field and inset shows the resistivity as a function of temperature at different applied magnetic field. (b) MR% in transverse magnetic field at different temperatures and inset shows the MR% in low field regime (≤ 1 Tesla). (c) Kohler's scaling plot at different temperatures.

Fig. 5: Magneto-conductivity as a function of transverse magnetic field at temperatures varies from 100K down to 2K fitted using (a) HLN equation in low field regime (up to ±1Tesla) and (b) HLN+$\beta H^2$+$\gamma H$ equation in entire applied field range.

Fig. 1

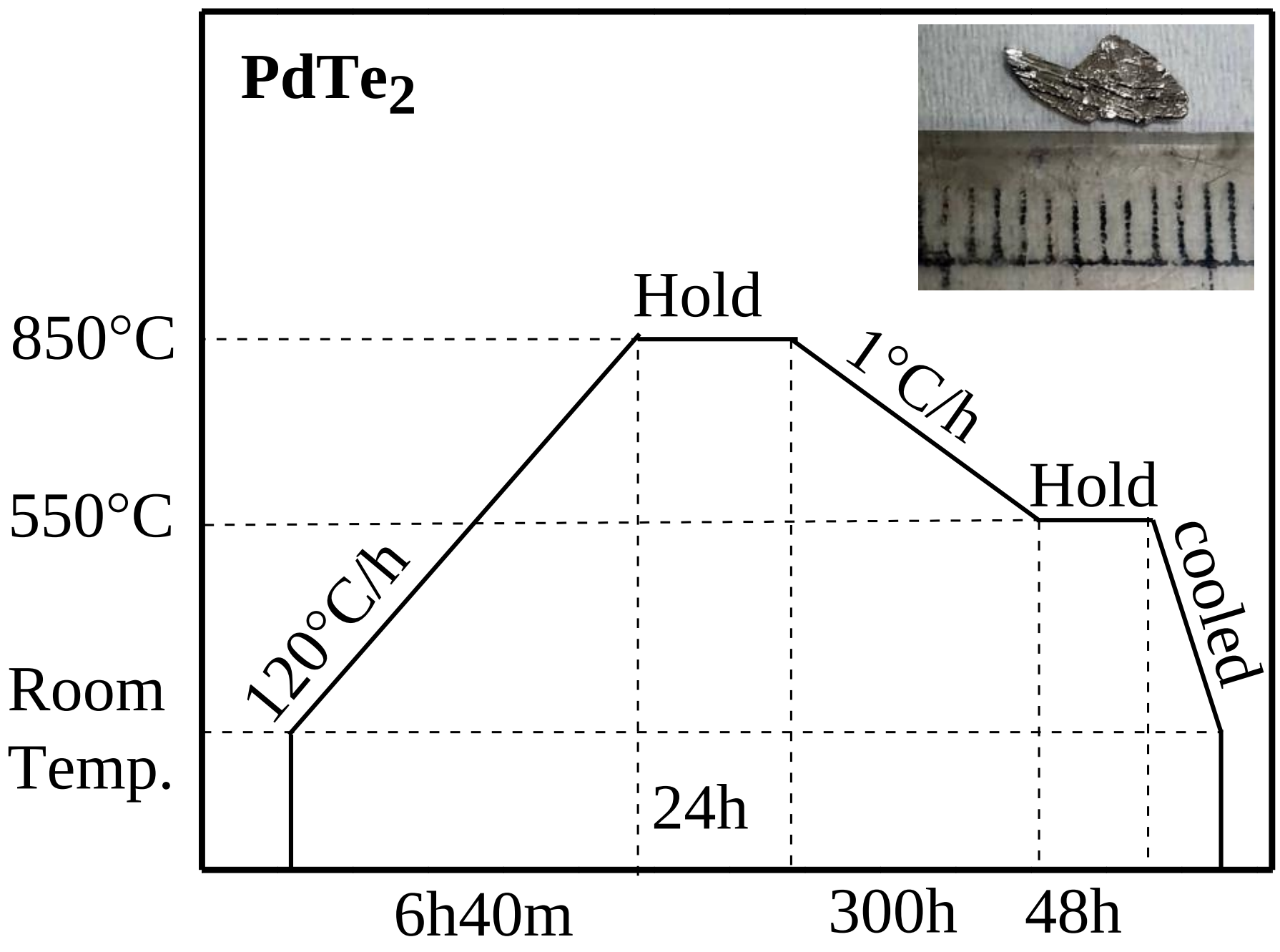

Fig. 2 (a)

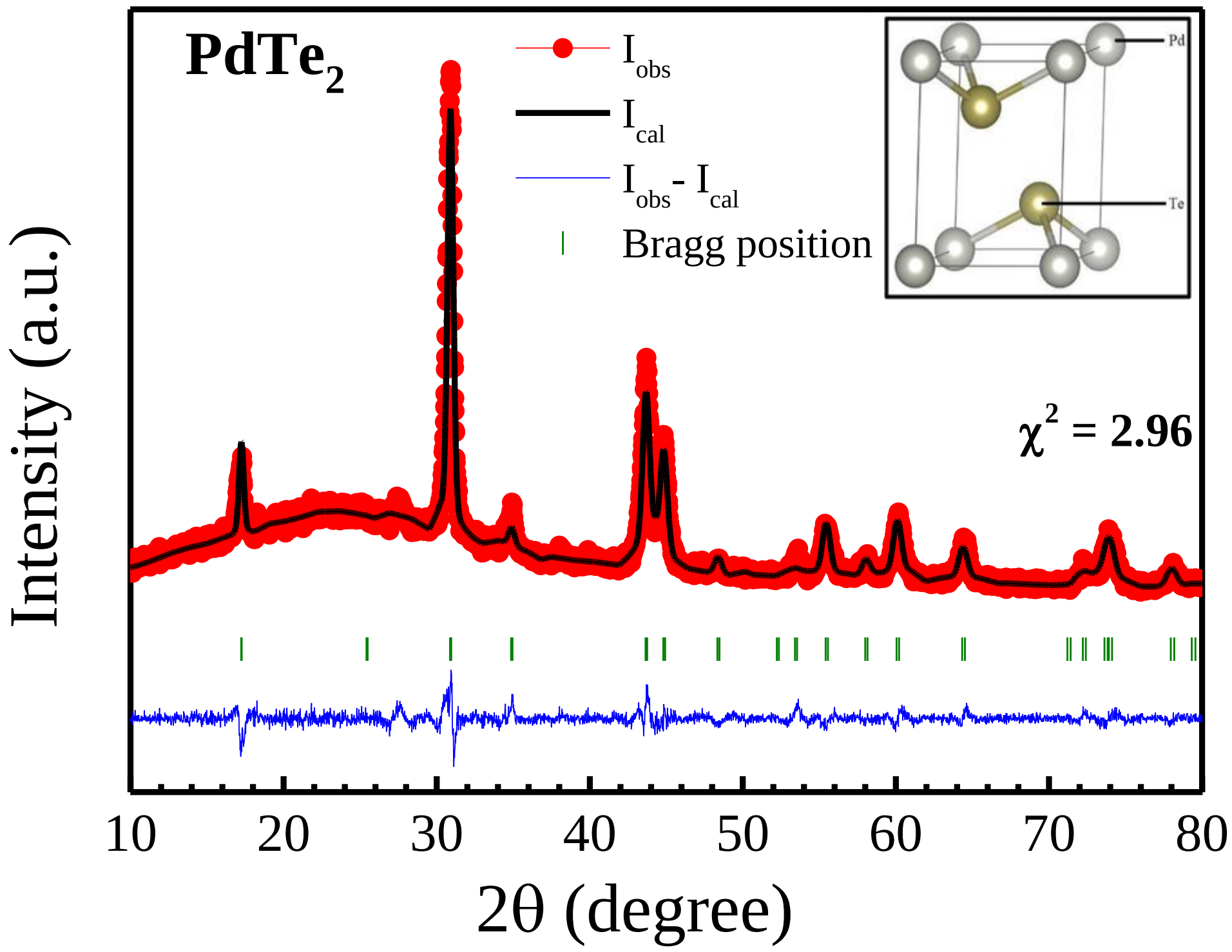

PdTe2
Iobs
Ical
Iobs - Ical
Bragg position
χ2 = 2.96
Intensity (a.u.)
10
20
30
40
50
60
70
80
2θ (degree)

Fig. 2 (b)

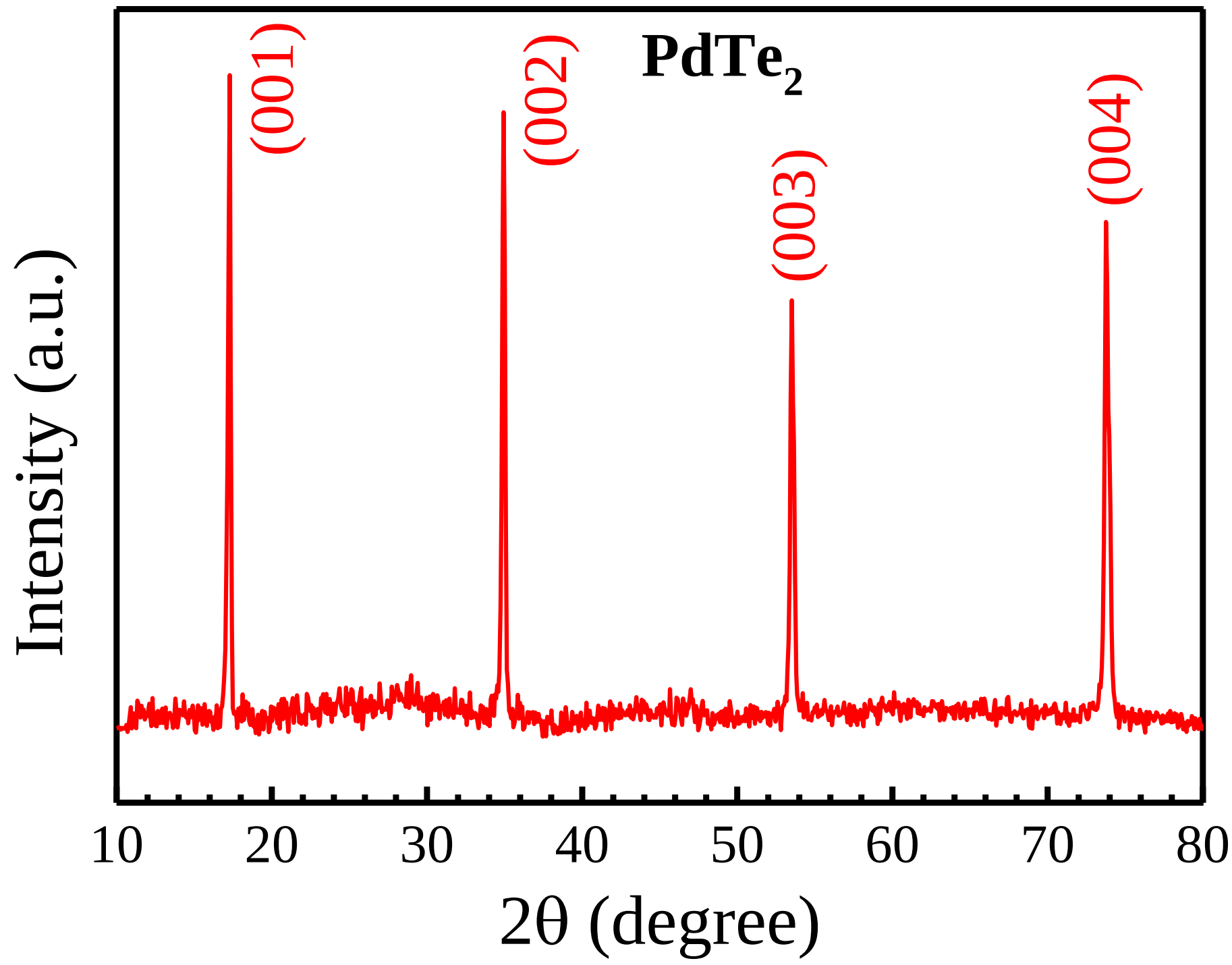

PdTe2
(001)
(002)
(003)
(004)
Intensity (a.u.)
10
20
30
40
50
60
70
80
2θ (degree)

Fig. 3

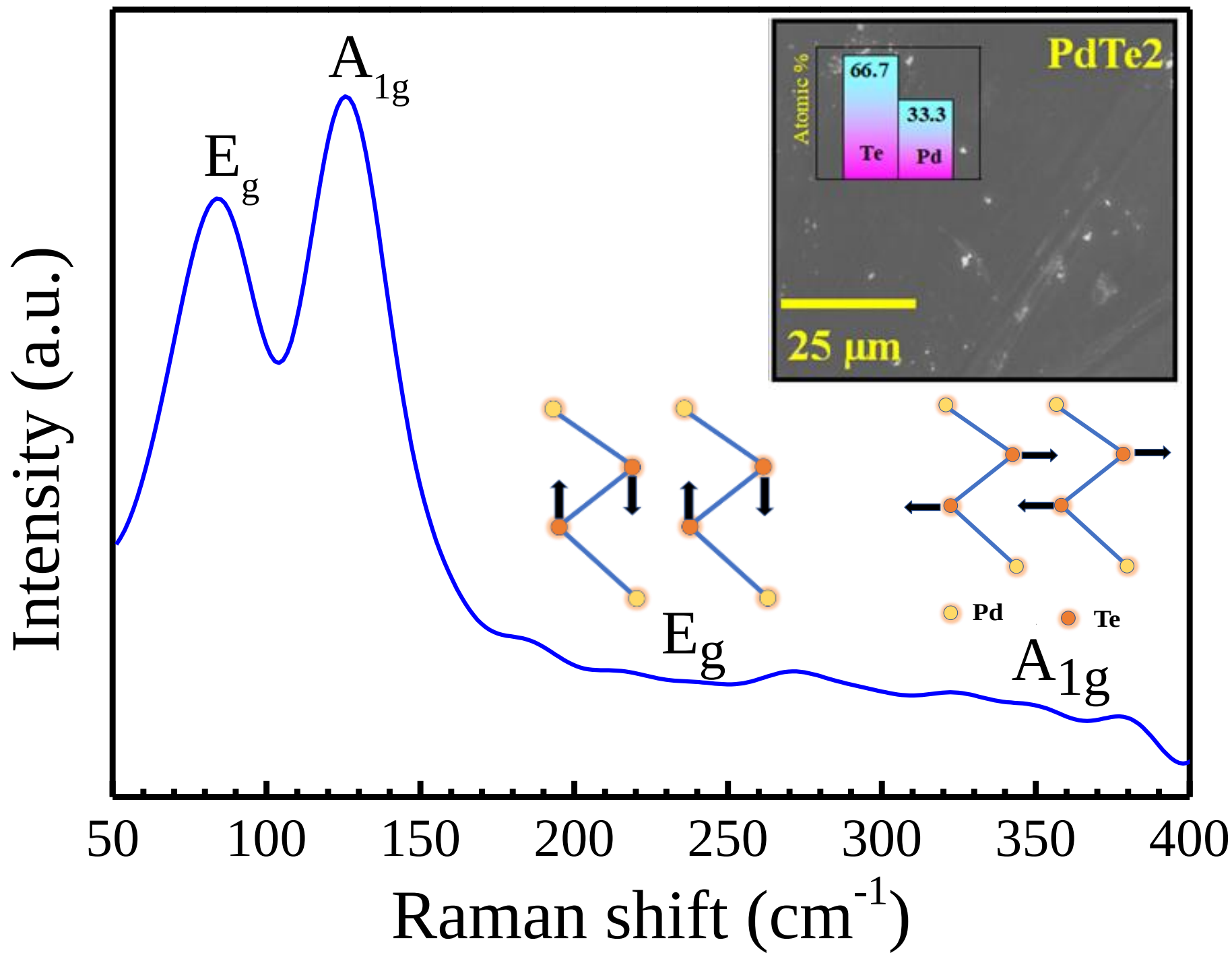

PdTe2
Atomic %
66.7
33.3
Te
Pd
25 μm
$E_g$
$A_{1g}$
$E_g$
$A_{1g}$
Pd
Te
Intensity (a.u.)
50
100
150
200
250
300
350
400
Raman shift ($cm^{-1}$)

Fig. 4 (a)

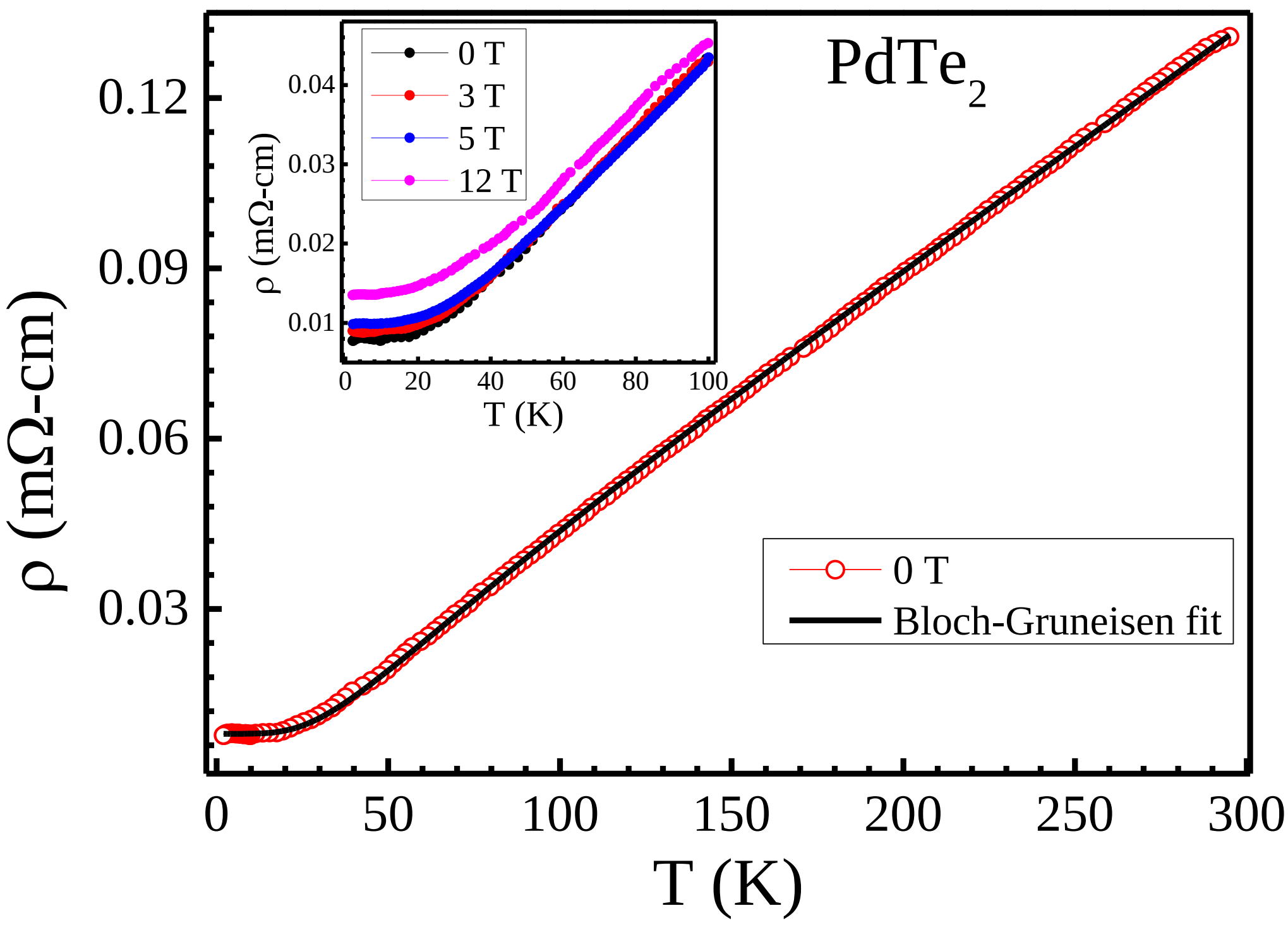

$PdTe_2$
0 T
3 T
5 T
12 T
0.04
0.03
0.02
0.01
ρ (mΩ-cm)
0
20
40
60
80
100
T (K)
0 T
Bloch-Gruneisen fit
0.12
0.09
0.06
0.03
ρ (mΩ-cm)
0
50
100
150
200
250
300
T (K)

Fig. 4 (b)

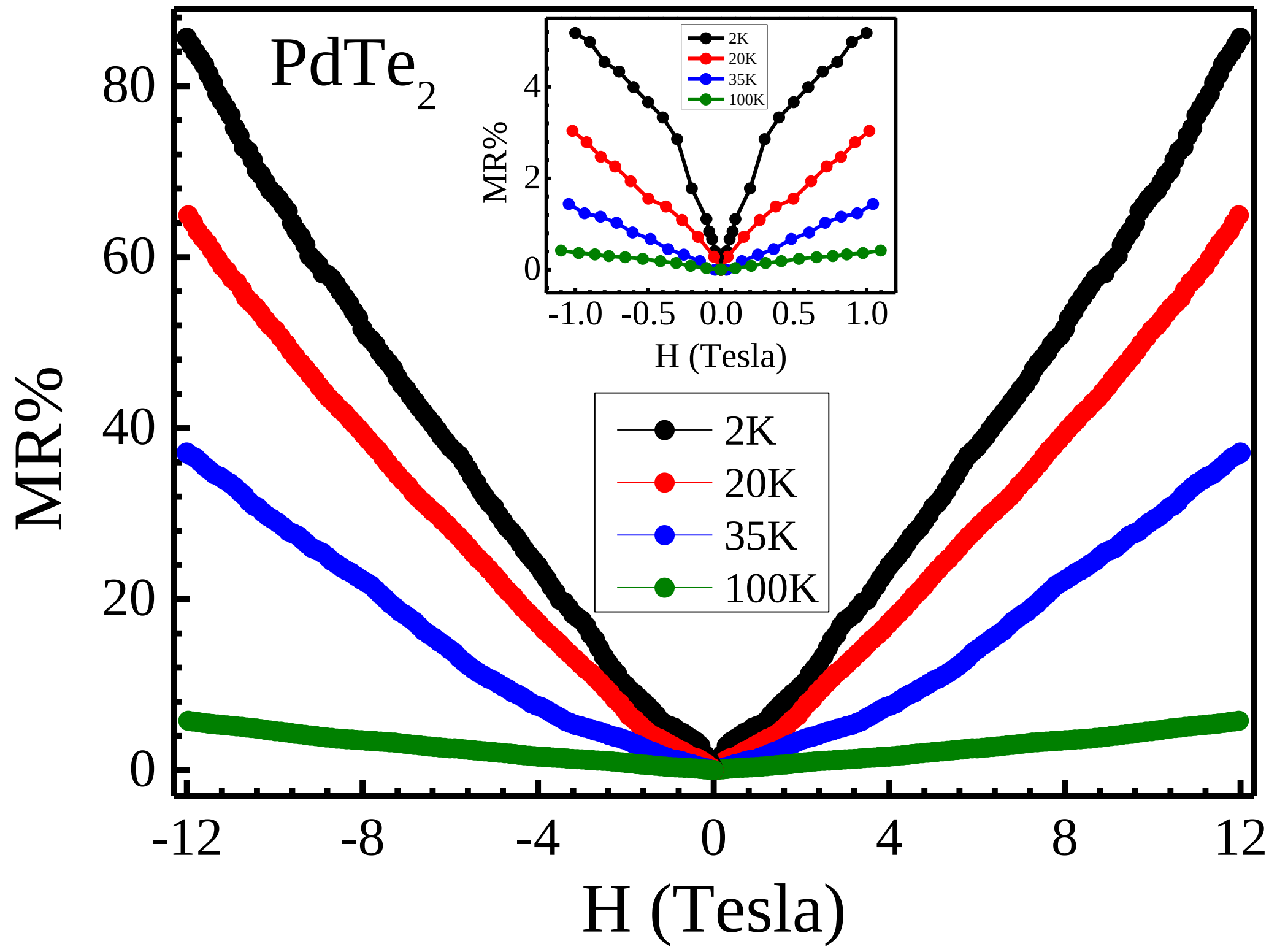
PdTe2
MR%
H (Tesla)
2K
20K
35K
100K
MR%
H (Tesla)
2K
20K
35K
100K

Fig. 4 (c)

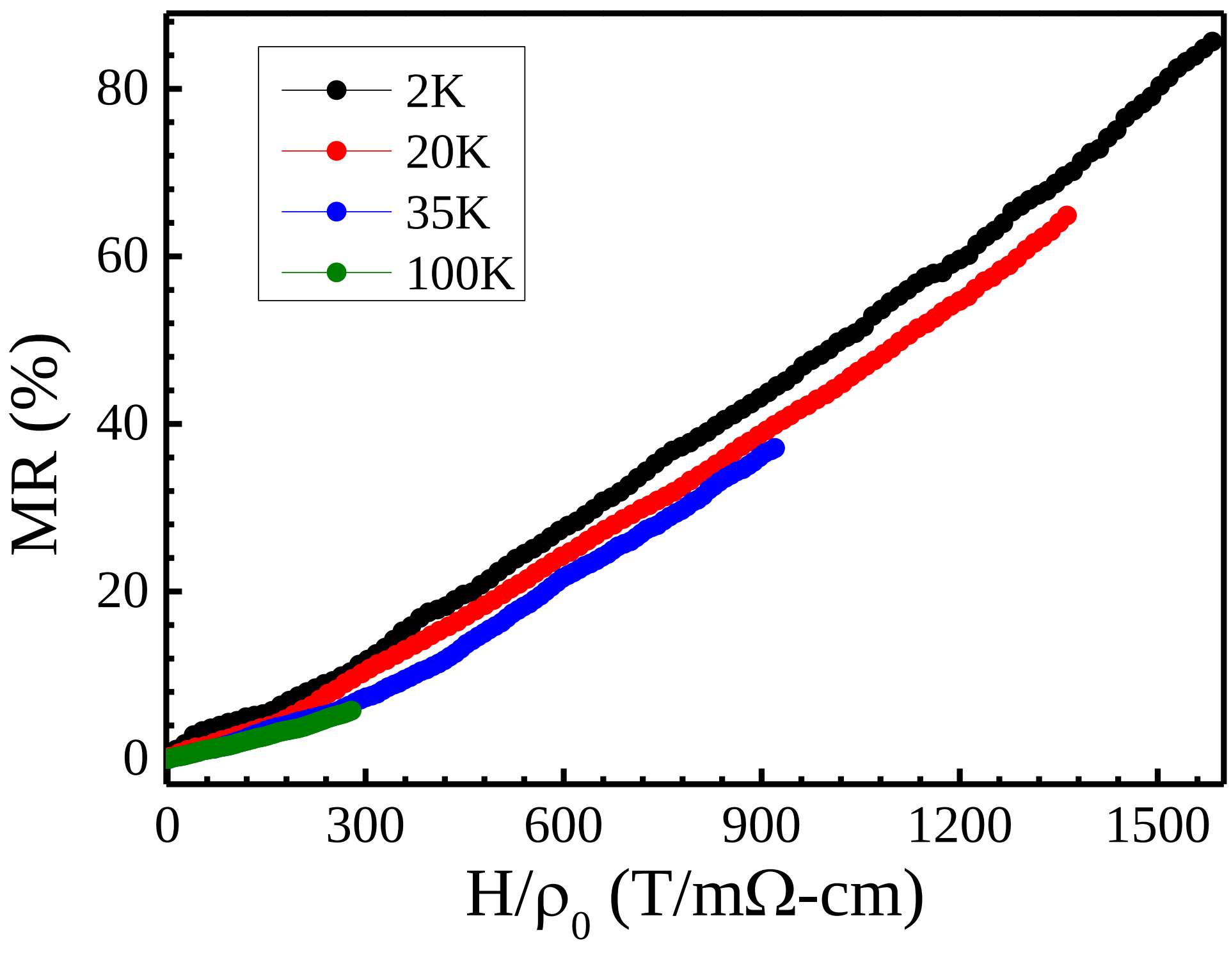
2K
20K
35K
100K
MR (%)
H/ρ0 (T/mΩ-cm)

Fig. 5 (a)

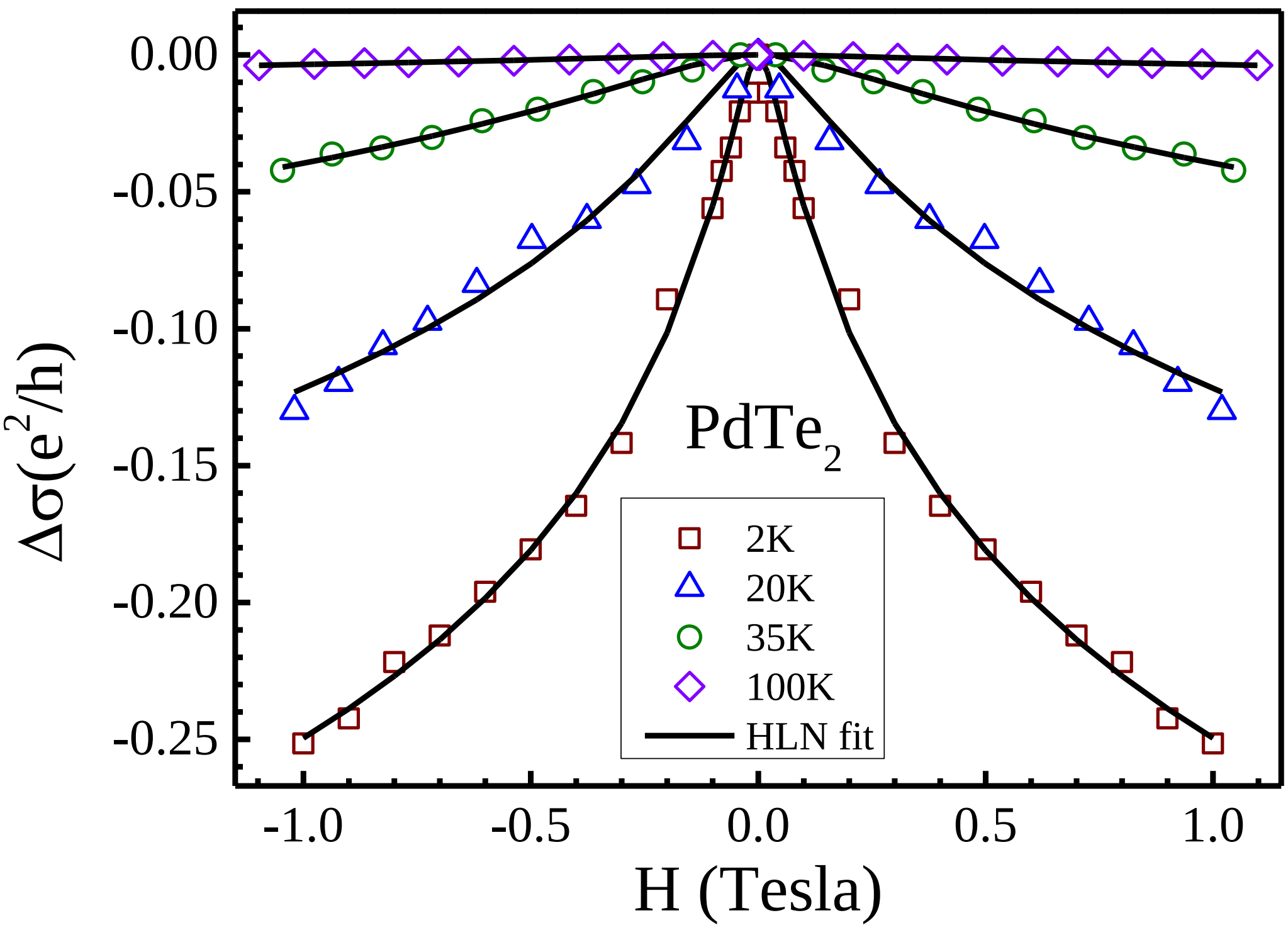

PdTe$_2$
2K
20K
35K
100K
HLN fit
$\Delta\sigma(e^2/h)$
H (Tesla)
0.00
-0.05
-0.10
-0.15
-0.20
-0.25
-1.0
-0.5
0.0
0.5
1.0

Fig. 5 (b)

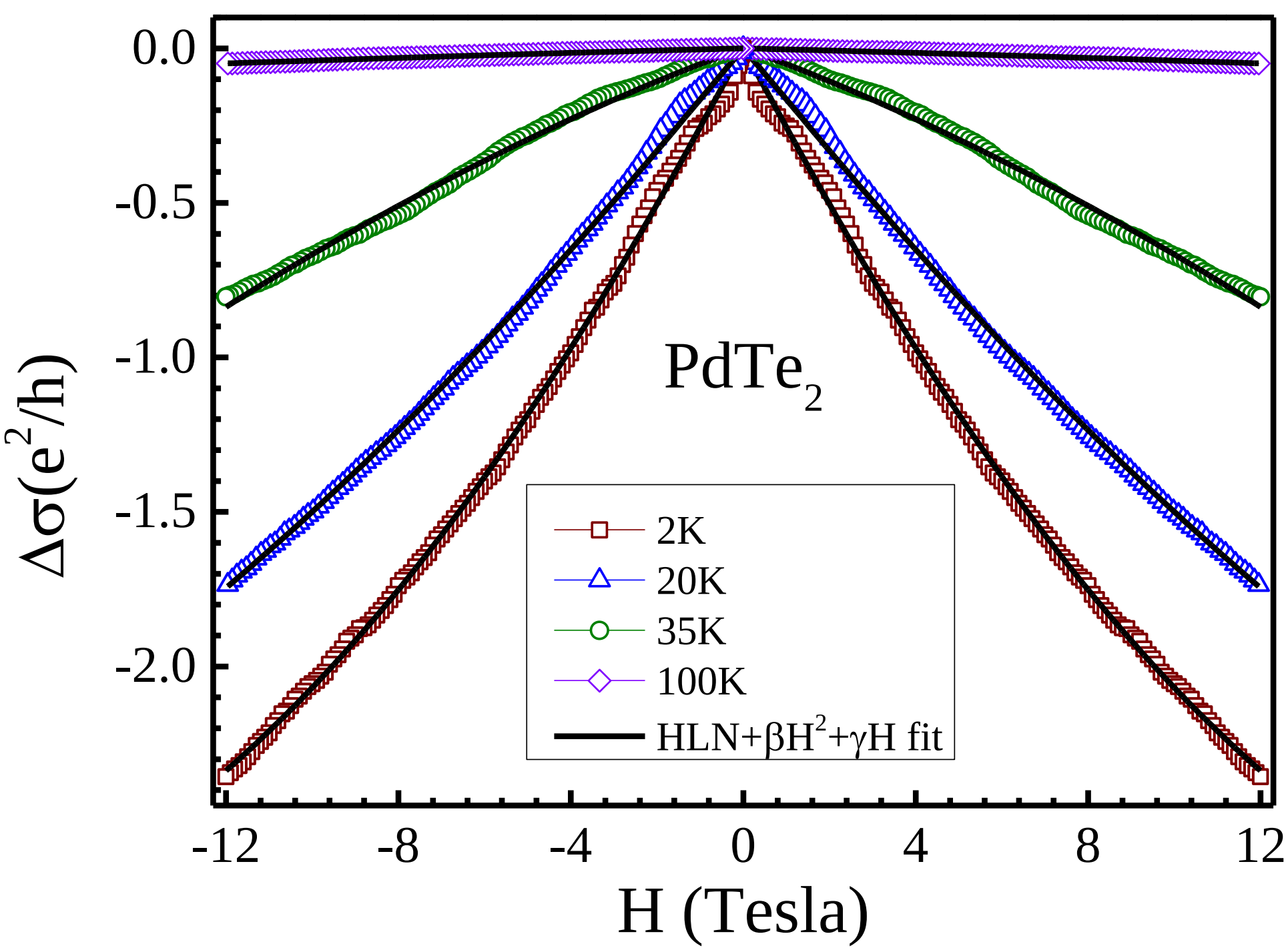

PdTe$_2$
2K
20K
35K
100K
HLN+$\beta H^2$+$\gamma H$ fit
$\Delta\sigma(e^2/h)$
H (Tesla)
0.0
-0.5
-1.0
-1.5
-2.0
-12
-8
-4
0
4
8
12